\documentclass[a4paper,11pt]{article}
\pdfoutput=1 % if your are submitting a pdflatex (i.e. if you have
\usepackage{jcappub} % for details on the use of the package, please
\usepackage[T1]{fontenc} % if needed
\usepackage{subfig}

\def\be{\begin{equation}}
\def\ee{\end{equation}}
\def\barr{\begin{array}}
\def\earr{\end{array}}
\def\bea{\begin{eqnarray}}
\def\eea{\end{eqnarray}}
\def\bfig{\begin{figure}}
\def\efig{\end{figure}}
\def\calH{{\cal H}}
\def\eqn#1{eq.\ (\ref{#1})}
\def\fig#1{Fig.\ (\ref{#1})}
\def\lt{\left}
\def\rt{\right}

\newcommand{\nn}{\nonumber}

\title{Cosmic viscosity as a remedy for tension between PLANCK and LSS data}
\author[a]{Sampurn Anand,}
\author[a]{Prakrut Chaubal,}
\author[a]{Arindam Mazumdar,}
\author[a]{Subhendra Mohanty}

\affiliation[a]{Physical Research Laboratory, Ahmedabad, 380009, India}
\emailAdd{sampurn@prl.res.in}
\emailAdd{prakrutchaubal@gmail.com}
\emailAdd{arindam@prl.res.in}
\emailAdd{mohanty@prl.res.in}

\abstract{
Measurements of $\sigma_8$ from large scale structure observations show a discordance
with the extrapolated $\sigma_8$ from Planck CMB parameters using $\Lambda$CDM cosmology.
Similar discordance is found in the value of $H_0$ and $\Omega_m$. In this 
paper, we show that the presence of viscosity, shear or bulk or combination of both,
can remove the above mentioned conflicts simultaneously.
This indicates that the data from Planck CMB observation and different LSS observations
prefer small but non-zero amount of viscosity in cold dark matter fluid.
}


\begin{document}
\maketitle
\flushbottom

%%%%%%%%%%%%%%%%%%%%%%%%%%%%%%%%%%%%%%%%%%%%%%%%%%%%%%%
\section{Introduction}
\label{sec:intro}
Over the past few decades, several observations have indicated that our Universe 
is dominated by dark components, namely, dark matter
(DM) and Dark energy (DE)~\cite{Zwicky:1933gu,Rubin:1970zza,Perlmutter:1996ds,Perlmutter:1998np,
Riess:1998cb,Hinshaw:2012aka,Ade:2015xua,Troxel:2017xyo}.
In the light of current observations from Cosmic Microwave Background (CMB) and
Large Scale Structure (LSS) observations, most favorable theoretical 
construct to understand the evolution of our Universe is provided by the Cold Dark Matter with 
cosmological constant, $\Lambda$, also referred as the {\it standard model of cosmology} ($\Lambda$CDM model), which is 
characterized by six parameters only. Predictions of $\Lambda$CDM cosmology have 
been seen successfully in the CMB observations. However, LSS observations
have shown some conflicts with it. In this paper we will address  these issues 
and will ameliorate these conflicts by introducing dissipative effects in the system.
 
 The standard six parameters of $\Lambda$CDM model are :  the
 ratio of density of cold dark matter and baryonic matter   
 to the critical density, $\Omega_{\rm cdm}^0$ and $\Omega_b^0$ respectively, evaluated today, the 
 acoustic scale $\Theta_{\rm MC}$, the amplitude ($A_s$) and the spectral index ($n_s$)
 of the primordial density perturbations and the optical depth to the epoch of reionization ($\tau_{\rm reion}$).
The value of Hubble parameter at current epoch $H_0$ is related to the acoustic scale, $\Theta_{\rm MC}$. As a 
consequence, one of them is considered as input model parameter leaving other as a derived quantity. 
These parameters are inferred from two different observations namely CMB and LSS. 
This type of indirect determination has mostly provided the value of $H_0$ lower than the 
direct measurement from type-IA supernova~\cite{Freedman:2000cf}. However, before the release of Planck data~\cite{Ade:2015xua}
the inferred value of $H_0$ from CMB was in agreement with that of the LSS observations.

The success of $\Lambda$CDM model lies in its capability of describing the observed quantities in large scales
and small scales in a single theoretical framework. The primary CMB anisotropies provide an estimate of the 
amplitude of the matter fluctuations at the last scattering surface. Given a cosmological model, these 
primary fluctuations can be extrapolated to provide an estimate of matter fluctuations at a later time
in the Universe. However, the above described framework for $\Lambda$CDM model predicts the value of $\sigma_8$,
r.m.s fluctuation of perturbations at 8$h^{-1}$ Mpc scale,
which is not agreement with other low-redshift observations of 
large-scale structure~\cite{Vikhlinin:2008ym,Macaulay:2013swa,Battye:2014qga,MacCrann:2014wfa,Aylor:2017haa,Raveri:2015maa,Lin:2017ikq}.
These disagreements in the value of $H_0$ and $\sigma_8$ inferred from CMB and LSS observations
are typically attributed  either to the signals of new physics or to the
systematic errors. The recent result from dark energy survey has also shown the 
similar tension in $\sigma_8-\Omega_m^0$ plane~\cite{Abbott:2017wau,Troxel:2017xyo}, which 
indicates some new physics might be responsible for this mismatch.

Several attempts have been made in this regard to address this discordance between CMB and LSS 
observations. It has been argued that the interaction between dark matter and dark energy~\cite{Pourtsidou:2016ico, Salvatelli:2014zta,Yang:2014gza} as well as dark matter and dark
radiation~\cite{Ko:2016uft,Ko:2016fcd,Ko:2017uyb}
can resolve this tension to some extent.
However, in most of the cases such models resolve one of the above mentioned tensions but fails 
to solve the other one. More importantly, interaction between the dark sectors can also modify the scale 
corresponding to matter radiation equality~\cite{Yang:2014gza} which might introduce greater 
problem than $\sigma_8$ mismatch. Some other attempts have been made by modifying the neutrino 
sector~\cite{Wyman:2013lza,Battye:2013xqa,Riemer-Sorensen:2013jsa}. Addition of massive sterile neutrino in the system
has been reported to reduce some tension in $\sigma_8-\Omega_m^0$ plane but not in
$H_0-\Omega_m^0$ plane~\cite{Wyman:2013lza,Battye:2013xqa}. Recently, it has been claimed
that quartessence models, where a single dark component mimics both dark matter and dark 
energy can reduce the tension in $\sigma_8 - \Omega_{m}^0$~\cite{Camera:2017tws}.

It has been discussed that the viscosity in CDM has the ability of reducing 
power on the small length scales~\cite{Blas:2015tla,Velten:2014xca,Thomas:2016iav}. The effect of bulk 
has been investigated extensively in~\cite{Velten:2014xca}. On the other hand, the  effect of shear
has also been investigated to some extent~\cite{Barbosa:2017ojt}. 
Attempts to quantify the dissipative effects in dark matter has been done in ref.~\cite{Kunz:2016yqy} from 
Baryon Acoustic Oscillation (BAO) data.
For recent review on this
topic, refer to~\cite{Brevik:2017msy}. The bulk viscosity suppresses the growth of structures 
by imparting a negative pressure against the gravitational collapse while the shear
viscosity reduces the amount of velocity perturbations which in turn stops the growth.
%Another distinction between bulk and shear viscosity is that the former acts homogeneously 
%and isotropically  while the later one breaks these symmetries. 
Therefore, on the small
length scales, where the homogeneity and isotropy are broken due to velocity gradients, 
effects of shear viscosity is expected to play crucial role. Although the physics of
these viscosities are different, we will show that there effect on large scale structure
is more or less similar. 

This paper is organized as follows: in section~\ref{sec:vis_theory}, we 
discuss the basic setup of viscous cosmology. This section is divided into two 
subsections \ref{sec:ptbn} and \ref{sec:growth}. In~\ref{sec:ptbn} we outline the 
cosmological perturbation theory and derive the perturbation equations in presence 
of the two viscosities. Further, in~\ref{sec:growth} we show the effect of these
viscosities on the growth of density perturbations which in turn effects the 
matter power spectrum. Having discussed the effects on the matter power spectrum, we move on
to perform Markov-Chain-Monte-Carlo (MCMC) analysis with data from different CMB and LSS
observations. In section~\ref{sec:s8-om}, we show that the inclusion of viscosities removes
the tension between the Planck and LSS data in  $\sigma_8-\Omega_m^0$ plane. Similarly,
in section \ref{sec:H0-om} we show the concordance between Planck and LSS
data in $H_0-\Omega_m^0$ plane due to viscosities. We also perform a joint analysis
of Planck and LSS data with viscosities to infer the parameters of 
viscous cosmology in section~\ref{sec:combined}. Finally, we conclude and discuss our
results in section \ref{sec:conc}.

%%%%%%%%%%%%%%%%%%%%%%%%%%%%%%%%%%%%%%%%%%%%%%%%%%%%%%%

%%%%%%%%%%%%%%%%%%%%%%%%%%%%%%%%%%%%%%%%%%%%%%%%%%%%%%%
% \section{Dissipative Effects}
% %%%%%%%%%%%%%%%%%%%%%%%%%%%%%%%%%%%%%%%%%%%%%%%%%%%%%%%
% \label{sec:dissipative}
% 
% 
% \subsection{Bulk-viscosity}
% Fundamental and effective. 
% \subsection{Shear-viscosity}
% Fundamental and effective.
%%%%%%%%%%%%%%%%%%%%%%%%%%%%%%%%%%%%%%%%%%%%%%%%%%%%%%%
\section{Effect of viscosity on large scale structures}
\label{sec:vis_theory}
%%%%%%%%%%%%%%%%%%%%%%%%%%%%%%%%%%%%%%%%%%%%%%%%%%%%%%%
Cosmological perturbations have been computed with the assumption 
of homogeneity and isotropy on large scale in presence of an ideal 
fluid~\cite{Bardeen:1980kt,Bond:1984fp,Kodama:1985bj,Ma:1995ey}.
% Bulk-viscosity can  be incorporated in the fluid without breaking the homogeneity
% and isotropy, but the presence of shear viscosity in the fluid 
% breaks these symmetries. 
Although violation of these symmetries might be
prominent on small length scales, their effect might not be substantial 
on large scales. The reason behind it is that velocity gradient on smaller scales 
becomes significant.
Therefore, we go beyond the perfect fluid approximation 
by considering dissipative effects in the system.
In this section we will describe the 
cosmological perturbation theory with non-ideal fluid in the presence of 
shear and bulk viscosities. 

The energy momentum tensor for the non-ideal fluid is given as~\cite{Weinberg:1972kfs}
  %%%%%
  \be
  T^{\mu\nu}_{vf} = \rho\,u^\mu\,u^\nu\, +\, (p + p_b)\,\Delta^{\mu\nu} +
  \pi^{\mu\nu}\, ,
  \label{eq:t-mu-nu-vf}
  \ee
  %%%%%
  where $\rho$ is the energy density and $p$ is the pressure in the rest frame of the fluid,
  $p_b = -\zeta\, \nabla_\mu\,u^\mu$ is the bulk viscous pressure with bulk viscosity $\zeta$.
  $\pi^{\mu\nu}$ is the shear-viscous tensor which takes the following form
  %%%%
  \be
  \pi^{\mu\nu} =  -2\eta\,\sigma^{\mu\nu}
  = -2\eta\,
  \lt[\frac{1}{2}
    \lt(
 \Delta^{\mu\alpha}\nabla_\alpha u^\nu + \Delta^{\nu\alpha}\nabla_\alpha u^\mu
    \rt) - \frac{1}{3} \Delta^{\mu\nu}\lt(\nabla_\alpha u^\alpha \rt)
    \rt]\, ,
  \label{eq:shear-tensor}
  \ee
 %%%%%
  with $\eta$ being the shear viscosity and
  $\Delta^{\mu\nu} = u^\mu\,u^\nu + g^{\mu\nu}$ being the projection operator
  which projects to the subspace orthogonal to the fluid velocity.
  It is evident that $\pi^\mu_\mu = 0 = u_\mu\,\pi^{\mu\nu} $. 
  Conservation of energy momentum, $\nabla_\nu\,T^{\mu\nu} = 0$, leads
  to the viscous fluid dynamic equations~\cite{Blas:2015tla} 
  \bea
  u^\mu\nabla_\mu\rho + (\rho + p)\nabla_\mu u^\mu - \zeta
  \lt(\nabla_\mu u^\mu\rt)^2 -2 \eta\sigma^{\mu\nu}\sigma_{\mu\nu} & = & 0
  \, ,\label{eq:energy-cons} \\ \nn \\
  (\rho + p + p_b) u^\mu\lt(\nabla_\mu u^\alpha\rt) +
  \Delta^{\alpha\mu}\nabla_\mu(p+p_b) +
  \Delta^\alpha_\nu \nabla_\mu\pi^{\mu\nu} & = & 0\, .
  \label{eq:momentum-cons}
  \eea

%=================================
  \subsection{Perturbation equations}
  \label{sec:ptbn}
%=================================
Perturbation in the matter field is related to the perturbation
in the metric through Einstein's equation and vice-versa.
In general, metric tensor $g_{\mu\nu}$ can have scalar, vector and 
tensor perturbations which are independent of each other in linear order. 
In this analysis we consider only
scalar perturbations in the conformal-Newtonian gauge given as
  \be
  ds^2 = a^2(\tau)\lt[-(1+ 2\,\psi(\tau, \vec x))\,d\tau^2 (1-2\phi(\tau, \vec x))\,dx_i\,dx^i\rt]\, ,
  \label{eq:ptrb-metric}
  \ee
  where $\psi(\tau, \vec x)$ and $\phi(\tau, \vec x)$ are space-time dependent functions.
We assume a spatially flat universe consisting of one species  of
viscous CDM along with cosmological constant($\Lambda$). 
Normalization of the flow field $u^\mu$ as $u_\mu\,u^\mu =-1$ allows us to express
it in terms of coordinate velocity $v^i$ and the metric perturbations as
%++++++++++++++++
\be
u^\mu = \frac{1}{a\sqrt{1\,+\,2\psi\, -(1 - 2\phi)\, v^2}}\, (1,\, v^i) ~~
\approx \frac{1}{a}\, (1-\psi, v^i ) + {\rm higher\,\, order\,\, terms}\, .
\ee
%++++++++++++++++ 
We parametrize the density and pressure in terms of isotropic background and
spatially varying small perturbations as 
\bea
\rho_m(\tau, \vec x)& =& \rho_m(\tau) + \delta\rho(\tau, \vec x)\, ,\nn\\
p(\tau, \vec x) &= & p(\tau) + \delta p(\tau, \vec x)\, ,
\eea
with $\delta\rho,\, \delta p \ll \rho_m$. The background field satisfies the following equations:
   \be
  \calH^2  = \lt(\frac{\dot a}{a}\rt)^2  = \frac{8\pi\,G}{3}\lt(\rho_{m} + \Lambda\rt) a^2\, ,
  \label{eq:hubble}
  \ee
  \be
  \dot \rho_{m} + 3\,\calH\,\lt(\rho_{m} + p_{m}\rt) = 0\, ,
  \label{eq:cont}
  \ee
  where
  ${\cal H} = \dot a/a $ is the Hubble parameter and dot denotes
  the derivative with respect to the conformal time $\tau$. For the analysis 
below, we consider perturbations, up to linear order, in the variables 
$\delta \rho$,  $\delta p$, $v^i$, $\phi$ and $\psi$. 

 In order to expand the fluid dynamic equations, we use the normalized density
  contrast $\delta = \delta\rho/\rho_m$ and the velocity divergence $\theta = \nabla_i\, v^i$. 
  Moreover, $\delta p$ is related to the density perturbations
 through $w$, the equation of state parameter, as
  \be
  w = \frac{p_{m}}{\rho_{m}}\,,~~~~
  c_s^2 = \frac{\delta\,p}{\delta\rho}\,, ~~~~
  c_{\rm ad}^2 = \frac{\dot p_{m}}{\dot \rho_{m}}
  = w - \frac{\dot w}{3 {\cal H}(1+w)}\, ,
  \label{eq:var1}
  \ee
where $c_s$ is the speed of sound in the medium and $c_{\rm ad}$ is 
the adiabatic sound speed. 

Inclusion of these perturbations in \eqn{eq:energy-cons} and \eqn{eq:momentum-cons}
leads to the dynamical equations which govern the evolution of cosmological perturbations.
In Fourier space these equations can be  
written as~\cite{Blas:2015tla},
  \be
  \dot  \delta = -(1+w)\lt(\theta - 3\dot \phi\rt)
  - 3\calH\lt(c_s^2 -w\rt)\,\delta \, ,
  \label{eq:delta1}
  \ee
  \be
  \dot \theta = -\calH\theta + k^2\,\psi +
  \lt(\frac{c_s^2}{1+w}\rt)\,k^2\delta + 3c_{\rm ad}^2\calH\theta -
  \frac{4}{3}k^2\,\frac{\eta}{(1+w)\,a\,\rho_{m}}\,\theta\, .
  \ee\label{eq:theta1}
For viscous matter we will evaluate the quantities defined in eq.(\ref{eq:var1}).\\
{\bf Equation of state, $w$ } :
 For baryonic matter and CDM, pressure $p_{m} = 0$. However, in presence of bulk viscosity the 
effective pressure of CDM is equal to the bulk pressure
   $p_{b} = -\zeta\, \nabla_\mu u^\mu = -3\,\zeta\, \calH/a \, .$
 Therefore, the equation of state parameter for CDM is
\be
    w = -\frac{3\,\zeta\,\calH}{a\,\rho_{\rm cdm}}
      = -\frac{\tilde \zeta\, a}{\Omega_{\rm cdm}\, \tilde\calH} \, ,
 \label{eq:eos}
\ee
where $\tilde \zeta = 8\pi G\,\zeta/\calH_0 $ is a dimensionless 
parameter and $\tilde\calH = \calH/\calH_0$. Throughout the manuscript  
`0' in the superscript or subscript denotes the value of the quantity evaluated today.
\\
 {\bf Sound speed, $c_s^2$} : Assuming a constant bulk viscosity, one can calculate
\be
 c_s^2 = -\frac{\zeta\,\theta}{a\,\rho_{\rm cdm}\,\delta} 
       =  \frac{w\,\theta}{3\calH\delta} 
       = -\lt(\frac{\tilde \zeta\,a}{\Omega_{\rm cdm}\, \tilde\calH}\rt)
          \lt(\frac{\theta}{3\calH\delta}\rt)\, .
  \label{eq:csq}
\ee
{\bf Adiabatic sound speed, $c_{\rm ad}^2$} :
Using eq.(\ref{eq:eos}) and performing a little bit of mathematical manipulation, we obtain
\be
   c_{ad}^2 = 2w\,\lt(1 - {\Omega_{\rm cdm}\over 4}\rt)\, .
\ee
In terms of quantities defined above, the evolution equation for $\delta$ of CDM (\eqn{eq:delta1})
takes the following form
\be 
  \dot\delta = -\lt(1 -\frac{\tilde\zeta\,a}{\Omega_{\rm cdm}\,\tilde\calH}\rt)
                   (\theta -3 \dot\phi)~
              + ~\lt(\frac{\tilde\zeta\, a}{\Omega_{\rm cdm}\, \tilde\calH}\rt)
   \theta ~ - ~\lt(\frac{3\,\,\calH\,\tilde\zeta\,a}{\Omega_{\rm cdm}}\rt)\,\delta \, 
\label{eq:delta}
\ee
and the evolution equation for $\theta$ of CDM (\eqn{eq:theta1}) becomes
  \be 
  \dot\theta = -\calH\,\theta + k^2\psi 
        -\frac{k^2\,a\,\theta}{3\,\calH\,(\Omega_{\rm cdm}\,\tilde\calH-\tilde\zeta\, a)}
               \lt(\tilde\zeta + \frac{4\tilde\eta}{3}\rt)
        - 6\,\calH\,\theta\lt(1-{\Omega_{\rm cdm}\over4}\rt)
               \lt(\frac{\tilde\zeta\,a}{\Omega_{\rm cdm}\,\tilde\calH}\rt)\, ,
\label{eq:theta}
  \ee
where $\tilde\eta = 8\pi\,G\eta/\calH_0$ is a dimensionless parameter.

The Poisson equation for viscous cosmology remain unchanged and given as
\bea
\nabla^2\phi = 3\calH\dot\phi + {a^2\over 2}\rho_m(\delta+ 2\psi)\, . 
\eea
The Euler equation in Fourier space gives 
\bea
\dot\phi = {3 a^2(p_b+\rho_m)\over k^2}\theta - \calH\psi + {\eta a\over k^2}(3\dot\theta - k^2\psi+3 \calH\theta)\, .
\eea

%====================
\subsection{Growth factor and effect on matter power spectrum}
\label{sec:growth}
%=====================
It is evident from eq.(\ref{eq:delta}) and eq.(\ref{eq:theta}) that the 
growth of the overdense region gets affected by shear as well as bulk viscosity. 
While bulk viscosity directly slows down the collapse of the overdense region, 
shear viscosity imparts similar effect through the velocity 
perturbation. Thus, it is important to investigate the effects of
these viscosities on the evolution of $\delta$ and $\theta$ which in turn effects
the matter power spectrum~\cite{Velten:2014xca,Barbosa:2017ojt}.
\begin{figure}[!tbp]
%\begin{center}
\hspace{-0.75cm}
\subfloat[\label{plot:shear-growth}]{
\includegraphics[width=2.2in,height=2.1in,angle=0]{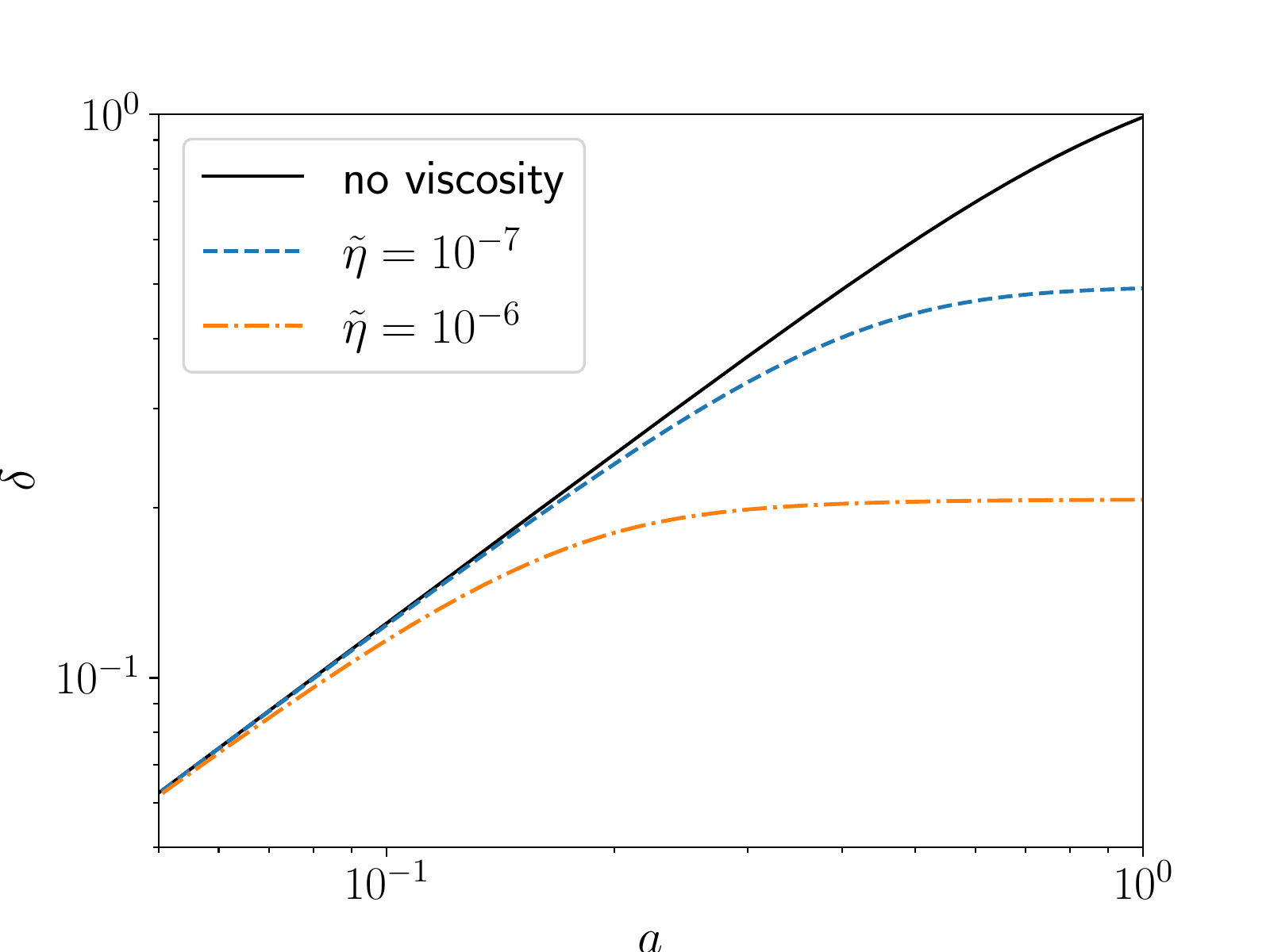}}
\hspace{-0.66cm}
\subfloat[\label{fig:bulk-growth}]{
\includegraphics[width=2.2in,height=2.1in,angle=0]{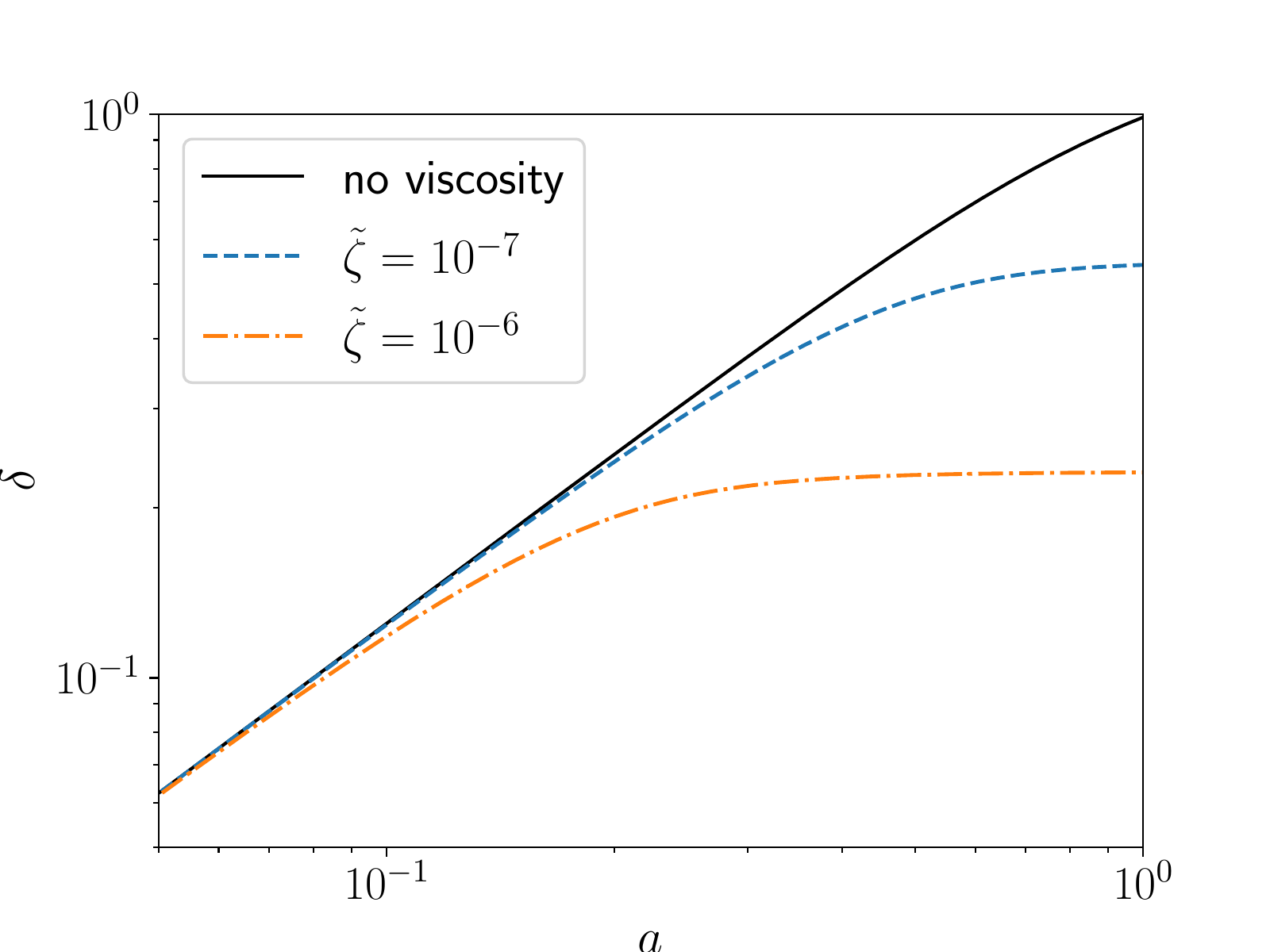}}
\hspace{-0.66cm}
\subfloat[\label{fig:delta-bulk-shear}]{
\includegraphics[width=2.2in,height=2.1in,angle=0]{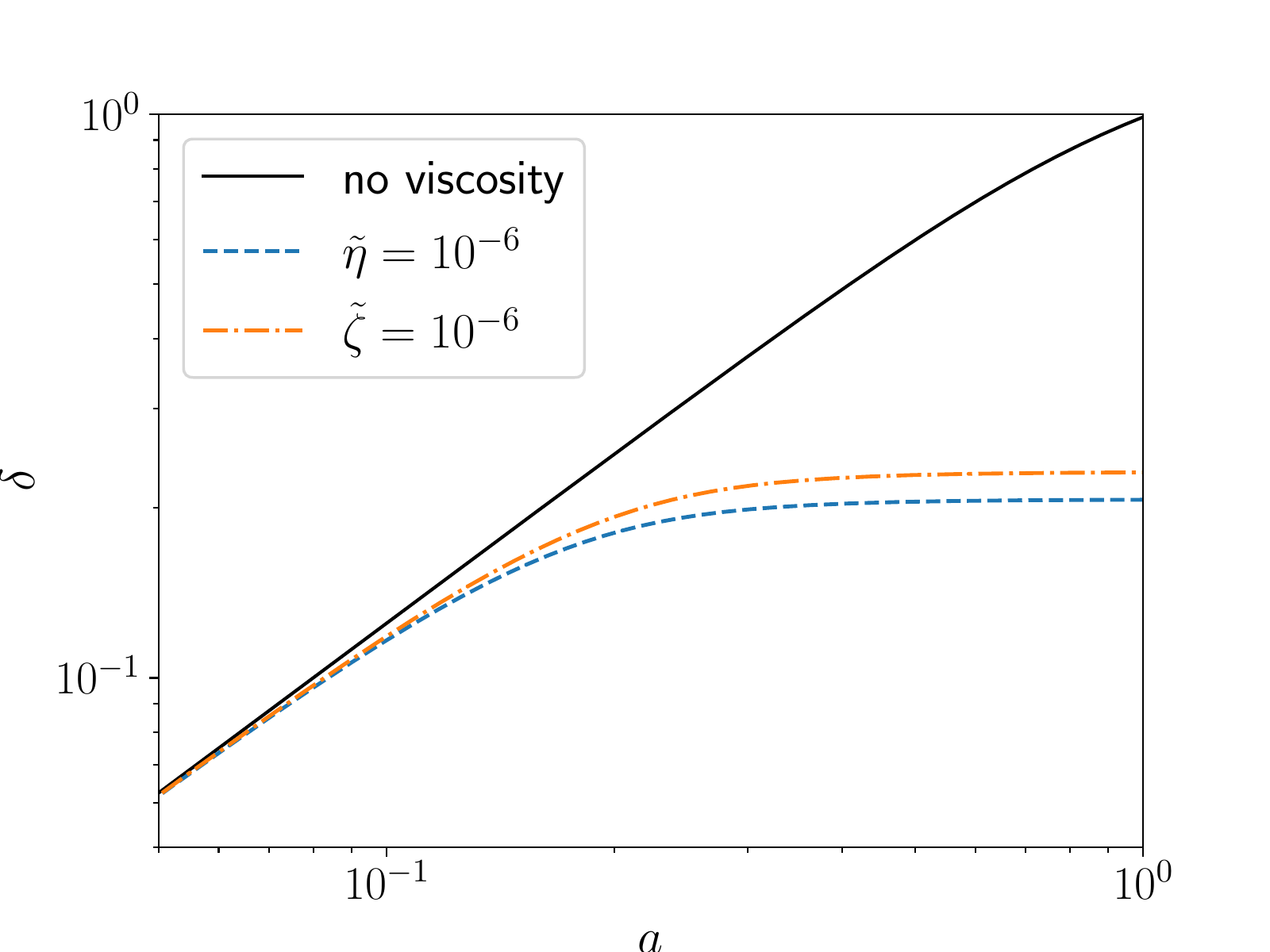}}
\caption{Effect of viscosity on the growth of linear over-density ($\delta$) as a
function of scale factor has been plotted for different values of~: (a) shear  (b) bulk viscosities. 
In figure~(c) we have compared the effect of bulk and shear viscosities. 
We have set $k\sim 3.4\,  h\,{\rm Mpc}^{-1}$ in all the three plots.}
   %\end{center}
\end{figure}

 We start by considering effect of shear viscosity on the evolution of 
perturbations. For this purpose, we set $\tilde \zeta =0$ and neglect 
$\dot\phi$. Therefore, \eqn{eq:delta} and \eqn{eq:theta}  can be rewritten as
\bea
 \dot \delta &= & - \theta \, ,\label{eq:delta-shear} \\
\dot \theta  & =&  -{\cal H}\theta + k^2 \psi -{4\,\tilde\eta\over 9}\,
 {a\,\calH_0\over\,\Omega_{\rm cdm}} {k^2\over\calH^2}\,\theta\, , 
\label{eq:theta-shear}
\eea
 with 
\be
 k^2 \psi = -{3\over 2}\Omega_{\rm cdm} {\cal H}^2\delta\, .
\ee
Note that there are two dissipative terms in the right hand side of  
eq.(\ref{eq:theta-shear}), namely the Hubble expansion and the shear viscosity term. If the shear term is
greater than the Hubble term then the evolution of 
$\theta$ is governed by shear along with the potential $\psi$. By comparing the
 first and last term of eq.(\ref{eq:theta-shear}), we can do an order of 
magnitude estimate for $\tilde\eta$ to influence the evolution of the velocity 
perturbation which ultimately influences the matter power spectrum. For $a=1$ case, $\tilde \eta$ turns out to be 
$\tilde\eta = {9\over 4}\lt(k\over{\cal H}_0\rt)^{-2}\Omega_{\rm cdm}^0$. 
For $k \sim 1 {\rm Mpc}^{-1}$ the value of shear viscosity is $\tilde \eta\sim {\cal O}(10^{-8})$.
 
\begin{figure}[!tbp]
\begin{center}
\subfloat[\label{fig:pow-shear}]{
\includegraphics[width=3.2in,height=2.6in,angle=0]{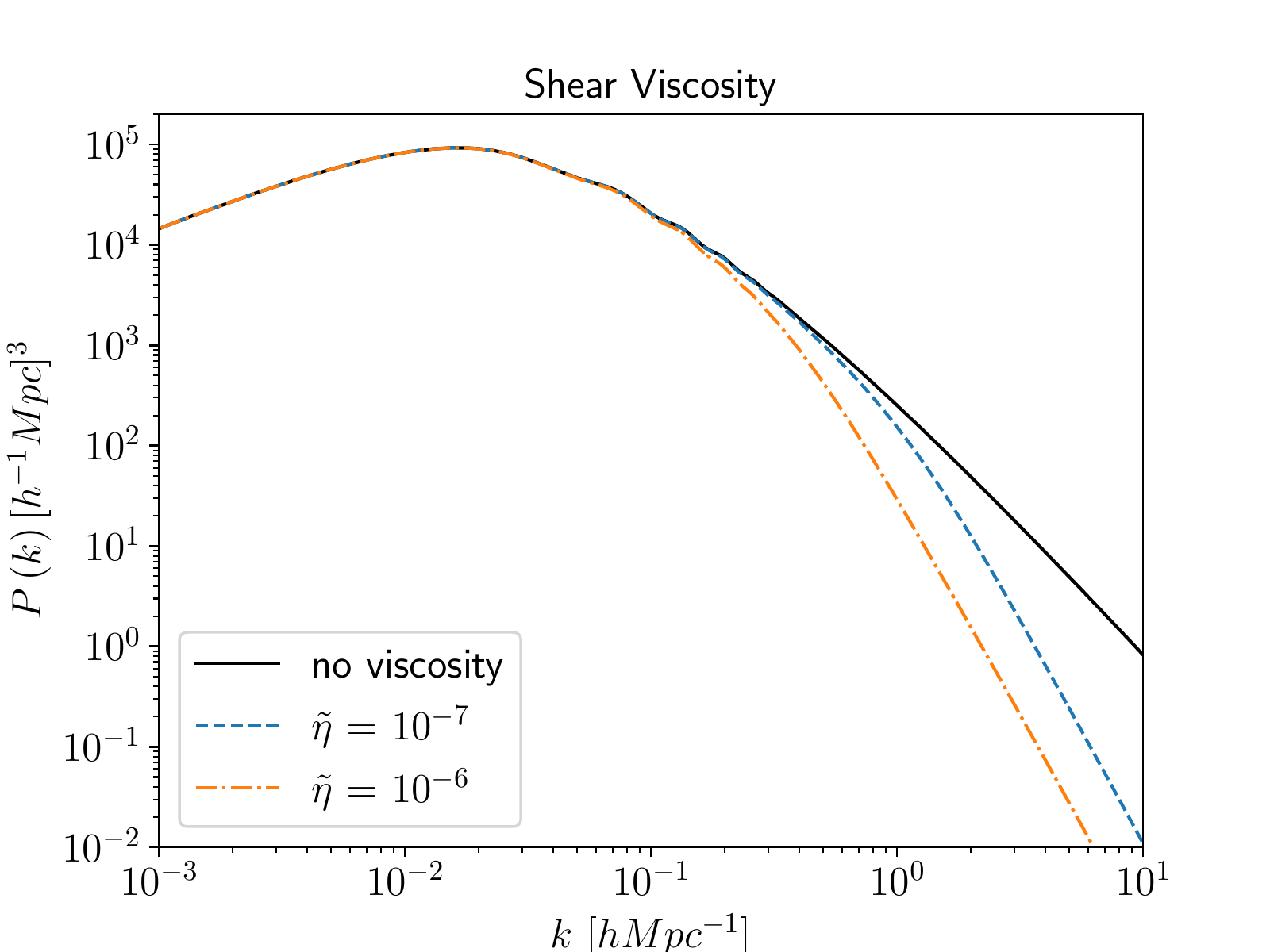}}~~
\hspace{-0.6cm}
\subfloat[\label{fig:pow-bulk}]{
\includegraphics[width=3.2in,height=2.6in,angle=0]{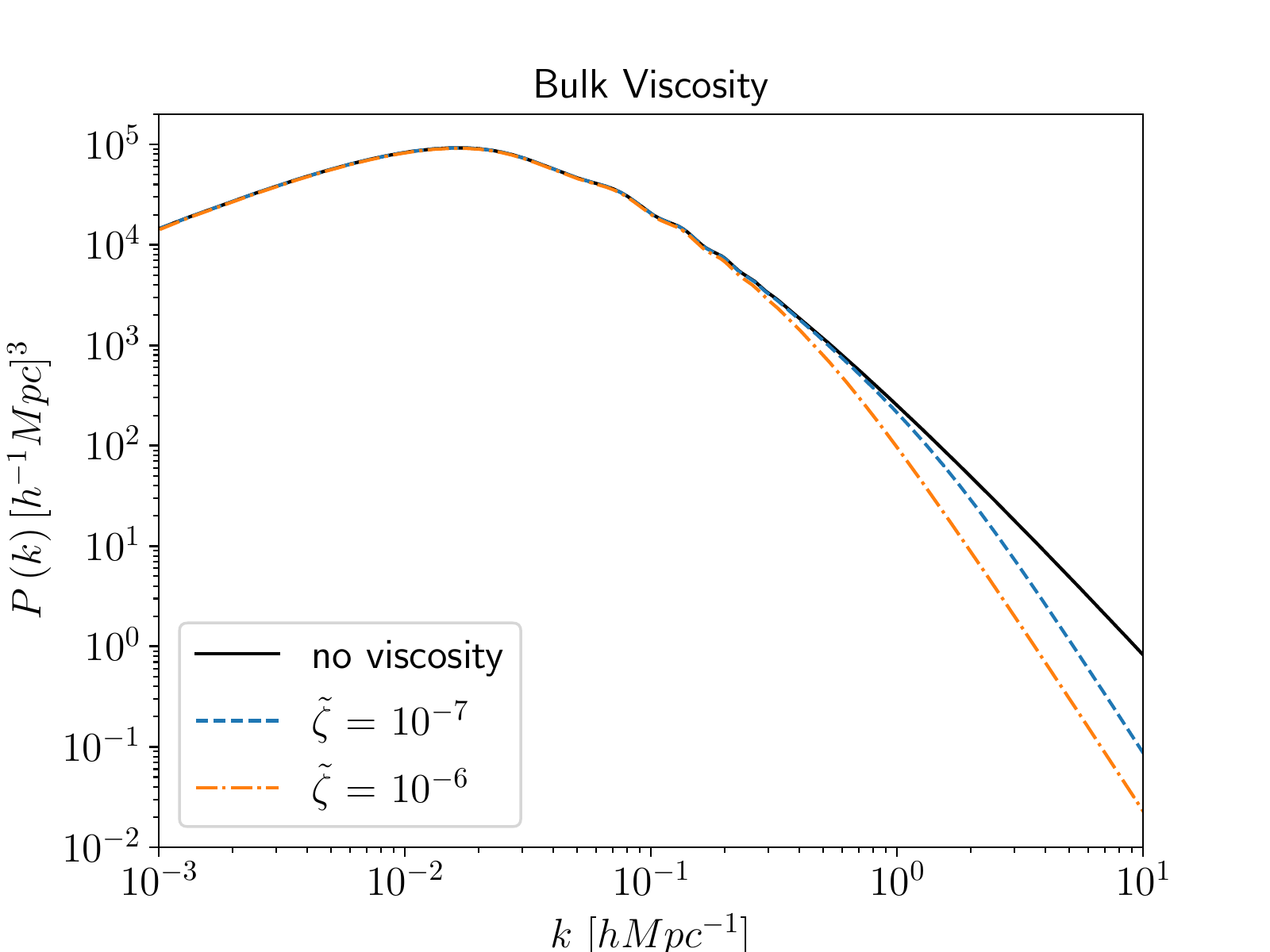}}
\caption{The effect of viscosities on the matter-power-spectrum is shown for
different values of (a) shear (b) bulk viscosities. It is evident that 
these viscosities play important role on large $k$.}
\end{center}
\end{figure}
 
 We combine the set of equations~\eqn{eq:delta-shear} and \eqn{eq:theta-shear} in a single second order differential equation
in terms of $a$ as,
\bea\label{eq:delta-prime}
\delta'' + \lt[{2\over a}+ {{\cal H}'\over {\cal H}}+ {4\over 9}{k^2\,\tilde\eta \,a\,\over \tilde\calH\,\Omega_{\rm cdm}^0}\rt]\delta ' 
-{3\over 2}{\Omega_{\rm cdm}^0\over a^3\,\tilde\calH^2}\delta =0 \, ,
\eea
where prime represents derivative with respect to the scale factor $a$.
Solution of this equation gives the growth of linear over density $\delta$ with 
$a$ which has been plotted in \fig{plot:shear-growth}. For solving this 
equation the initial value of $\delta$ at $a=10^{-3}$ has been set to
${2\over 5}{a\over \Omega_m^0}$ and the value of $\delta'$ is ${2\over 5}{a\over \Omega_m^0}$.
We see in \fig{plot:shear-growth} that the growth of linear overdensity
gets suppressed at late time. 

\begin{figure}[!tbp]
\begin{center}
\subfloat[\label{fig:rho_for_bulk}]{
\includegraphics[width=3.0in,height=2.2in,angle=0]{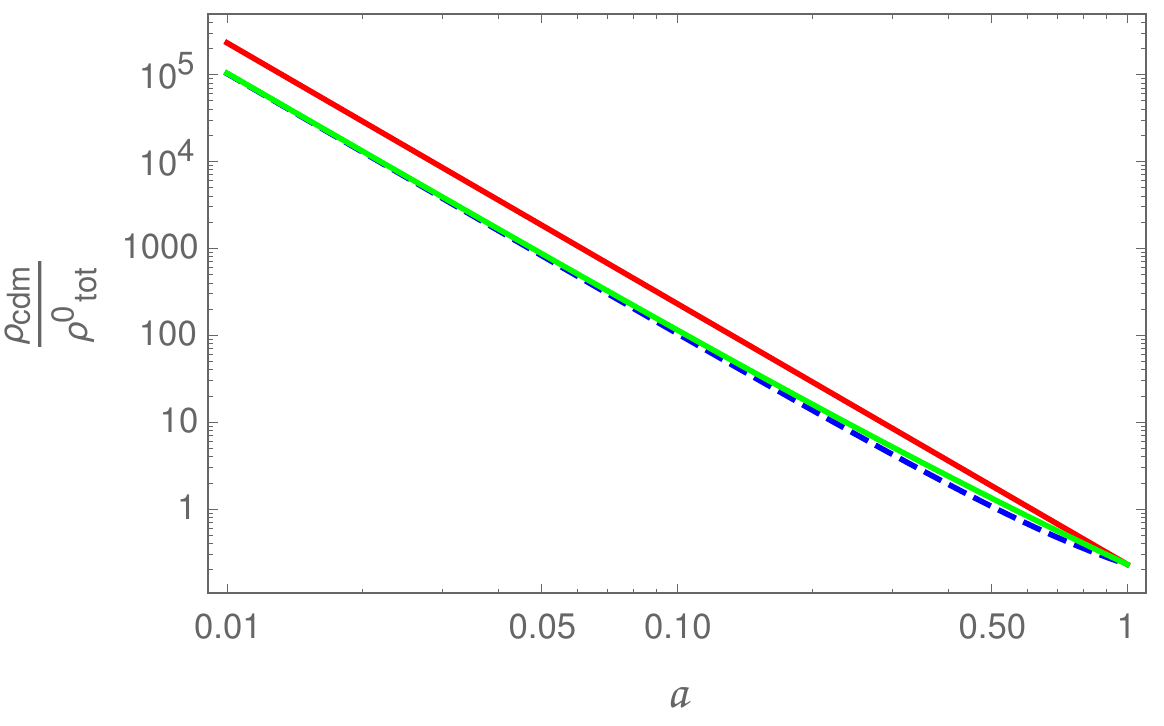}}~~~~
\hspace{-0.6cm}
\caption{ Change in the evolution of $\rho_{\rm cdm}$ in the presence of 
bulk viscosity. The blue-dashed line shows the numerical solution of
 \eqn{eq:bulk-cont} 
for $\tilde\zeta\sim 0.1$. The value of $\zeta$ has been taken too large for demonstration 
purpose. The green-solid line shows the fitting function given in \eqn{rho_bulk_fit} with $\beta=0.558$. 
The red-solid line represents standard  $\rho_{\rm cdm} = {\rho^0_{\rm cdm}\over a^3}$.}
\end{center}
\end{figure}

On the other hand, effect of bulk viscosity appears in two different ways. 
First, it modifies the  background evolution of cold dark matter and second, 
it changes the perturbation-equation for $\delta$ as well as $\theta$. Since 
bulk viscosity of dark matter changes the equation of state (see \eqn{eq:eos}),
 the evolution $\rho_{\rm cdm}$ with $a$ also gets modified which is depicted 
through the continuity equation (\eqn{eq:cont}) as
 \bea\label{eq:bulk-cont}
\rho_{\rm cdm}' + {3\rho_{\rm cdm}\over a}\lt(1-{\tilde\zeta\rho^0_{\rm tot}\over\rho_{\rm cdm}}
\lt[{\Omega^0_b\over a^3}+\Omega_\Lambda + {\rho^0_{\rm cdm}\over\rho_{\rm tot}}\rt]^{1/2} \rt) = 0\, .
 \eea
We solve this equation numerically and fit the solution 
(see \fig{fig:rho_for_bulk}), for numerical work, with a function of the following 
form 
 \bea\label{rho_bulk_fit}
\rho_{\rm cdm}(a) = {\alpha} {\rho^0_{\rm cdm}\over a^3} + {\beta} {\rho^0_{\rm cdm}\over a^2}\, ,
 \eea
where normalization at $a=1$ ensures that $\alpha = 1-\beta$. We have verified that
this form fits well even for large range of $\tilde\zeta$.
This form of $\rho_{\rm cdm}(a)$ will be used in numerical solution using CLASS~\cite{Lesgourgues:2011re,Blas:2011rf} 
later in the paper. The value of $\beta$ for $\tilde\zeta = 10^{-6}$ turns out to be $6.18\times 10^{-6}$.

To consider the effect of bulk on the perturbation equations,
we set $\tilde\eta = 0$ in equations (\eqn{eq:delta} and \eqn
{eq:theta}) and solve them numerically. The solution is plotted and shown in 
\fig{fig:bulk-growth}. Therefore, we can see that bulk viscosity has similar
effect as that of the shear viscosity. However, the effect of shear viscosity is slightly 
more than that of the bulk viscosity on the growth of delta as shown in \fig{fig:delta-bulk-shear}. 

The suppression of $\delta$ at late time due to viscosities shows its effect on matter 
power spectrum.  We have seen in \eqn{eq:theta} that the bulk and shear viscosities come 
into the equation multiplied with a $k^2$ factor. Therefore on small length scales
their effects become prominent.
This is expected as the velocity gradients are more effective on small
length scales resulting in large viscosity and hence suppressing the growth at 
those scales. Consequently, one would expect that the shear viscosity may 
influence $\sigma_8$.
To get the matter power-spectrum we have used publicly 
available CLASS code~\cite{Lesgourgues:2011re,Blas:2011rf}. 
We have not used non-linear halo-fit for evolution of $\delta$ at large $k$, since 
non-linear evolution of viscous dark-matter is beyond the scope of this paper. We will 
use this power spectrum to get the value of $\sigma_8$ which corresponds to $k\sim 0.78~h~{\rm MPc}^{-1}$,
the scale which is expected not to be effected by non-linearities.  
The power-spectrum has been plotted in \fig{fig:pow-shear} and \fig{fig:pow-bulk} where we can see
as expected the larger $k$ modes of $\delta$ gets more suppressed. In these figures
we have extended the linear analysis beyond $k=1~h~{\rm MPc}^{-1}$ only for the 
purpose of demonstration.

%
%Now we will see the effect of bulk-viscosity on the growth of over-density similarly like the shear viscosity.
%
%
%%%%%%%%%%%%%%%%%%%%%%%%%%%%%%%%%%%%%%%%%%%%%%%%%%%%%%%%%%%%%
\section{Resolving $\sigma_8$-$\Omega_m$ tension}
\label{sec:s8-om}
%%%%%%%%%%%%%%%%%%%%%%%%%%%%%%%%%%%%%%%%%%%%%%%%%%%%%%%%%%%%%
\begin{figure}
\hspace{-0.8cm}
\subfloat[\label{fig:eta-bestfit}]
{ \includegraphics[scale=0.40]{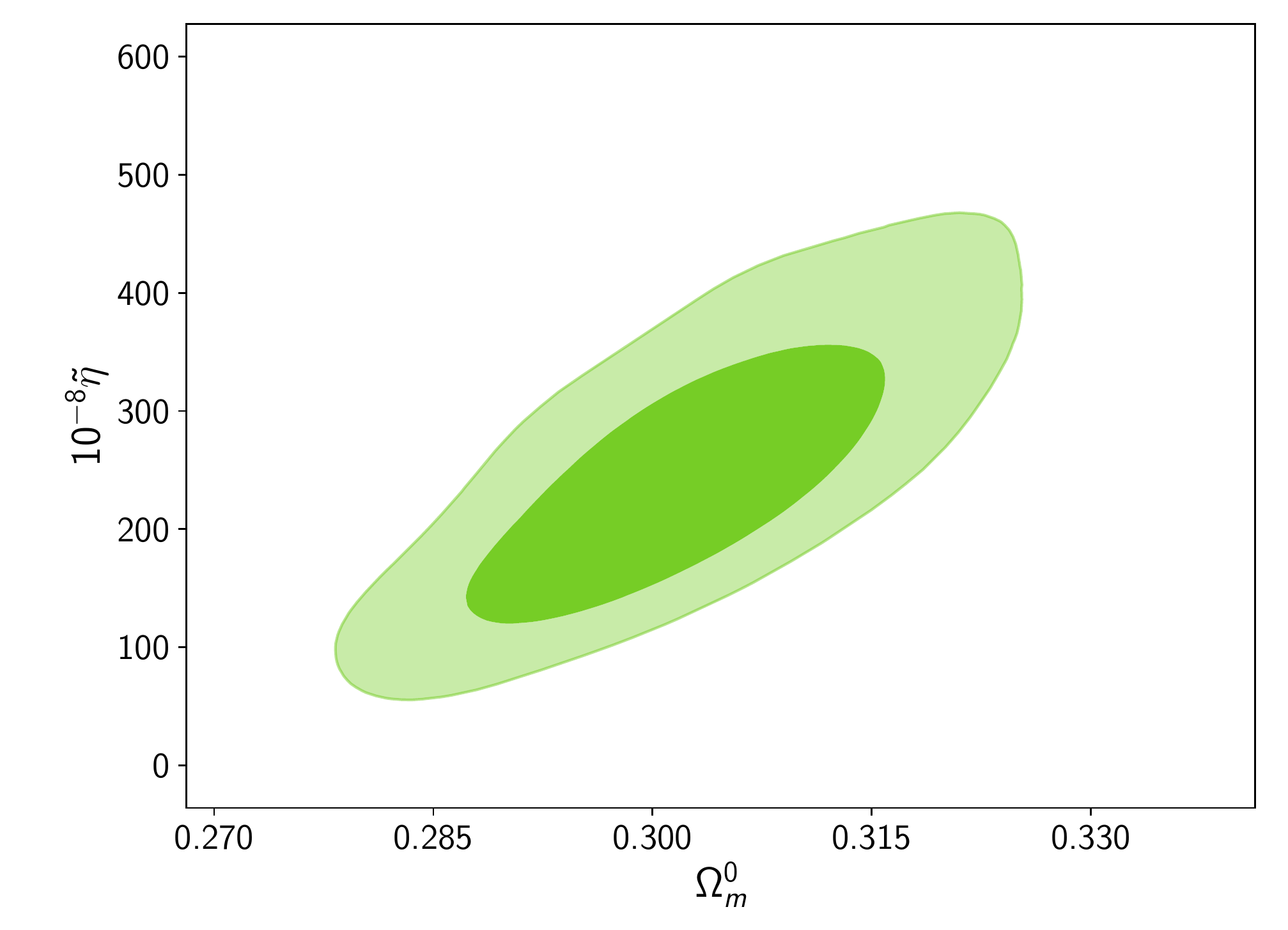} }
\hspace{-0.4cm}
\subfloat[\label{fig:shear-s8-om}]
{\includegraphics[scale=0.39]{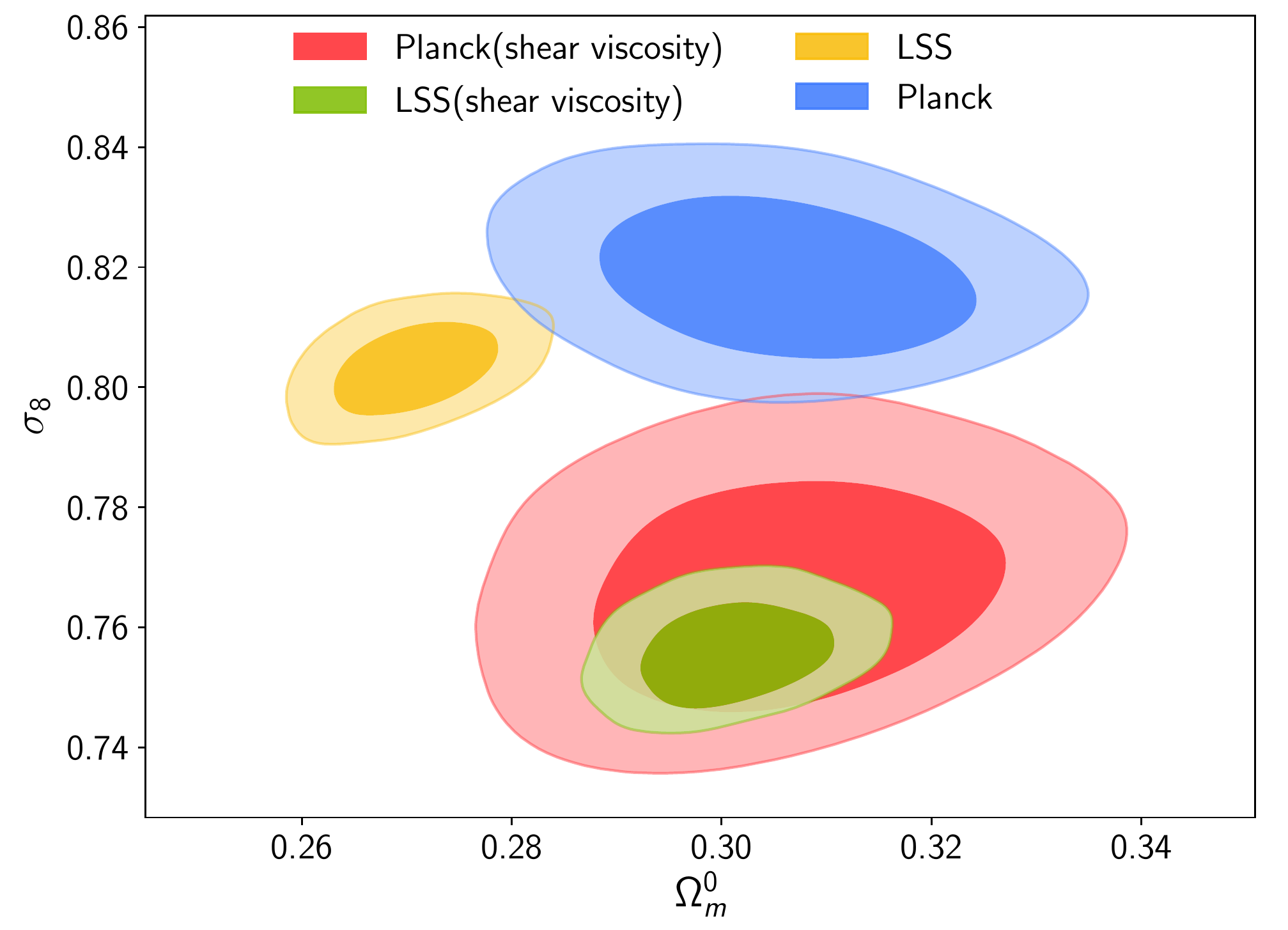}}
  \caption{(a) The best-fitted range of $\tilde\eta$ with 1-$\sigma$ and 2-$\sigma$
  contour for LSS data (Planck SZ + lensing, BAO-BOSS, SPT and CFHTLens) is shown. The central value of best fit is $ 2.30 \times 10^{-6} $. (b) We show the 
  Planck (high-$\ell$ + low -$\ell$) and LSS fitted region of $\sigma_8-\Omega_m^0$ which clearly shows the discordance. The discordance
  ends when the best-fit value of shear viscosity is used.
  }\label{fig:s8-om_m}
\end{figure}

\begin{figure}
\hspace{-.8cm}
\subfloat[\label{fig:zeta-bestfit}]
 {\includegraphics[scale=0.40]{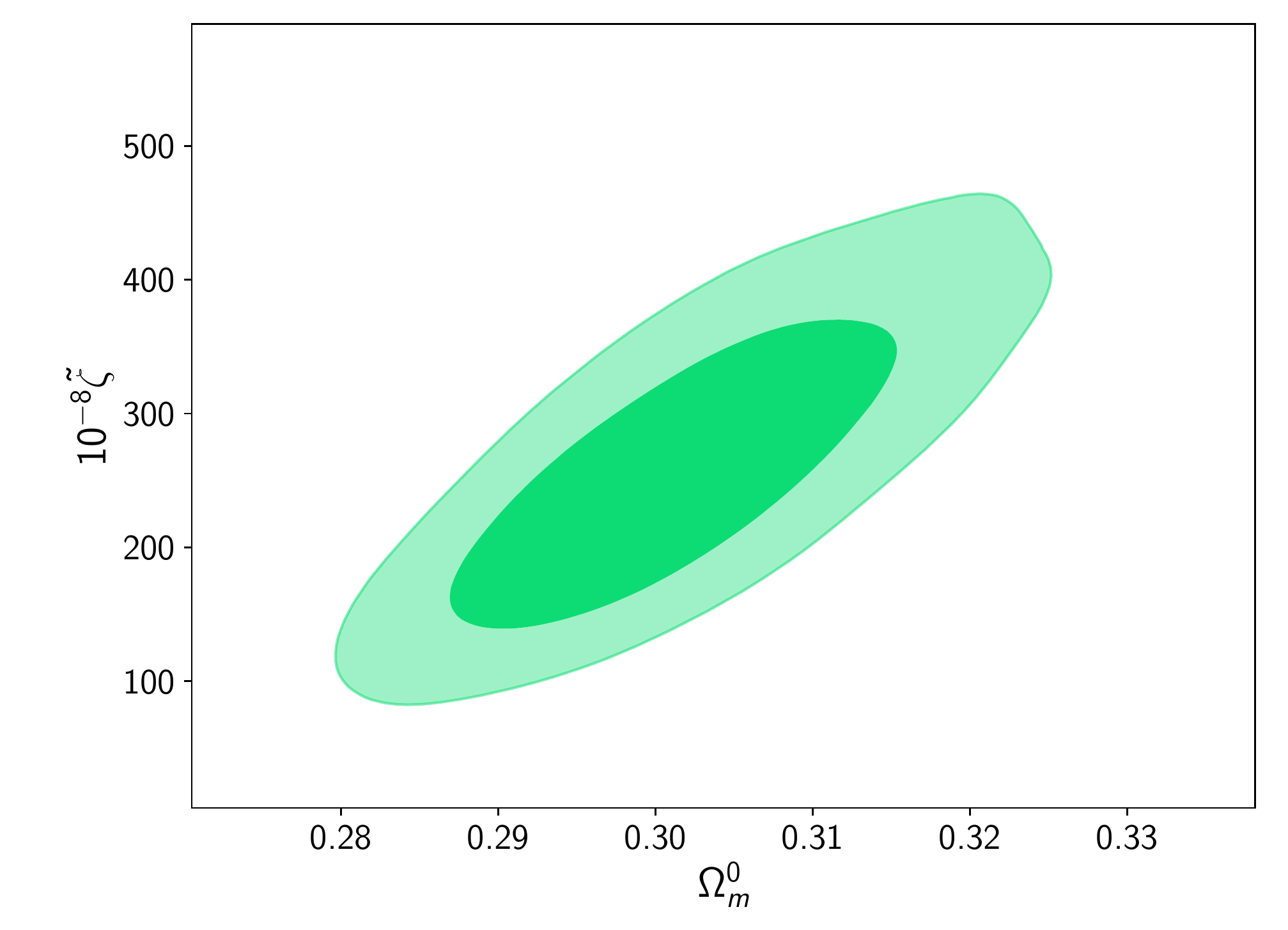}}
\hspace{-0.2cm}
 \subfloat[\label{fig:bulk-s8-om}]
 {\includegraphics[scale=0.39]{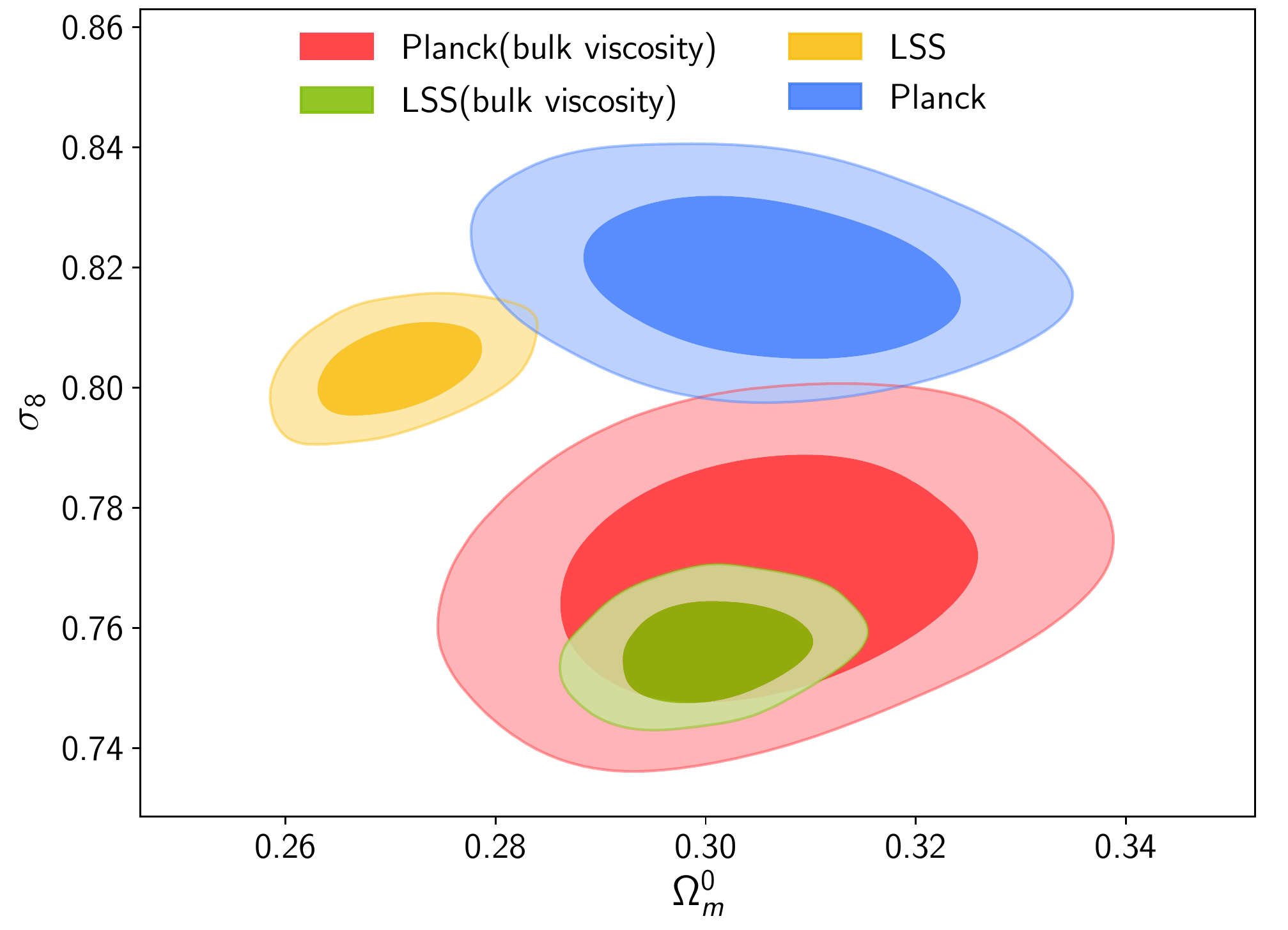}}
 \caption{(a) The best-fitted range of $\tilde\zeta$ with 1-$\sigma$ and 2-$\sigma$
  contour for LSS data  (Planck SZ + lensing, BAO-BOSS, SPT and CFHTLens)  is shown. The central value of best fit is $ 2.60 \times 10^{-6} $. (b) We show the 
  Planck (high-$\ell$ + low -$\ell$) and LSS fitted region of $\sigma_8-\Omega_m^0$ which clearly shows the discordance. The discordance
  ends when the best-fit value of bulk viscosity is used.
  }
\end{figure}
In this section we will show that there exist some tension between the LSS observations
and Planck CMB observation in $\sigma_8-\Omega_m^0$ plane. We will also demonstrate that
small but non-zero amount of viscosity in cold-dark matter can remove this 
tension.

In order to quantify our analysis we proceed in four steps. 
First, we  find the best-fit values of $\sigma_8$ and other cosmological parameters without viscosity
from Planck high-$\ell$ data and low-$\ell$ data (hereafter Planck data) using MCMC analysis~\cite{Audren:2012wb}. 
Throughout our analyses, we have considered massless neutrinos which changes the values of the parameters
slightly compared to the values obtained in ref.~\cite{Ade:2015xua}. 
Since $A_s$ and $n_s$ have the same origin in the early universe, we fix the 
priors on these quantities from the Planck parameter estimation which gives   
$\ln(10^{10}\times A_s)= 3.072\pm 0.027 $ and $n_s= 0.9681\pm 0.0058$.

In the next step, we find the best-fit values of $\sigma_8$ and other cosmological parameters without viscosity
from LSS data which include
Planck SZ survey~\cite{Ade:2013lmv}, Planck lensing survey~\cite{Ade:2013tyw}, 
Baryon Acoustic Oscillation data from BOSS~\cite{Anderson:2013zyy,Font-Ribera:2013wce},
South Pole Telescope (SPT)~\cite{Schaffer:2011mz,vanEngelen:2012va} and CFHTLens~\cite{Kilbinger:2012qz,Heymans:2013fya}
(hereafter LSS data).
%These data have one sigma overlap with each other~\cite{Battye:2014qga}.
 We keep $\tau_{\rm reion}$ fixed at $0.070$, since $\tau_{\rm reion}$ does not have much effect on LSS. 
These LSS surveys altogether indicate a value of $\sigma_8$ to be $0.8034_{-0.0098}^{+0.0104}$ at 2-$\sigma$ level
whereas Planck CMB observations predicts it to be $0.8186_{-0.0216}^{+0.0180}$ at 2-$\sigma$ level.
Therefore, there exits a mismatch between these two observations which is evident in $\sigma_8-\Omega_m$
plane shown in \fig{fig:shear-s8-om} and \fig{fig:bulk-s8-om}. 

We proceed to the next step which is to obtain the best-fit value for the viscosity parameters
$\tilde\eta$ and $\tilde\zeta$. In this step we keep
$A_s$ and $n_s$ prior as obtained from the analysis of Planck data.
The best-fit value for $ \tilde{ \eta } $ turns out to be $( 2.30\pm 0.58) \times 10^{-6}$ at 1-$\sigma$ level,
as shown in \fig{fig:eta-bestfit} and the best-fit value of $\tilde\zeta$ turns out to be 
$ (2.60\pm 0.78) \times 10^{-6}$ at 1-$\sigma$ level (\fig{fig:zeta-bestfit}).

Further, we set the values of viscosity parameters $\tilde\eta$ and $\tilde\zeta$
to their best-fit values obtained in the previous step. With these values we perform MCMC analysis
for Planck data to obtain the statistical estimates of standard cosmological parameters 
$ \{\Omega_b , \Omega_{cdm}, A_s, n_s, \Theta_{\rm MC}, \tau_{\rm reion} \}$ and the 
derived parameters $H_0$ and $\sigma_8$.
Finally, we perform similar analysis with LSS data by setting $\tilde\eta$ and $\tilde\zeta$ to their 
best-fit values. For this last step we keep the values 
of $A_s$ and $n_s$ obtained from the Planck as their prior.

As we have discussed in previous section, the viscosity in the cold dark matter reduces 
the power on small scales in matter power spectrum. As a consequence
$\sigma_8$ shifts downward in $\sigma_8 -\Omega_m$ fitting plane for
both the LSS and Planck CMB data. 
The result with viscosity shows a clear overlap of entire $1~\sigma$ region for $\sigma_8{\rm -}\omega_m$ which 
had earlier a $2~\sigma$ discordance (\fig{fig:shear-s8-om} and \fig{fig:bulk-s8-om}).

% Then, we follow method used by , 
% assuming the cosmological model determined by the Planck data is the correct cosmological model, 
% and set the priors of cosmological parameters : $\theta_{MC} = 1.04131 \pm 0.00063 $ and $n_s = 0.9603 \pm 0.0073$
% consistent with 1-$\sigma$ values determined with the Planck data to avoid over-fitting of the model. 
% We also fix the optical depth to re-ionization to $\tau_{reion} =  0.078$ as it has no effect on the 
% LSS data. With these priors set, we perform another MCMC analysis by varying $ \{ \Omega_b h^2, \Omega_b h^2, A_s \} $ 
% parameters with the LSS data. \fig{fig:s8-om} shows the results of these two analyses along with those of the similar analyses
% performed with the standard $\Lambda CDM$ cosmological model on the $\sigma_8$-$\Omega_m$ plane.

% \begin{figure*}
%  \includegraphics[scale=0.8]{figs/sig8-om_m.pdf}
% \end{figure*}
\section{Resolving $H_0$-$\Omega_m$ tension}
\label{sec:H0-om}
%=======================================================
\begin{figure}[!htbp]
\hspace{-0.7cm}
\subfloat[\label{fig:H0-om-bulk}]
{ \includegraphics[scale=0.40]{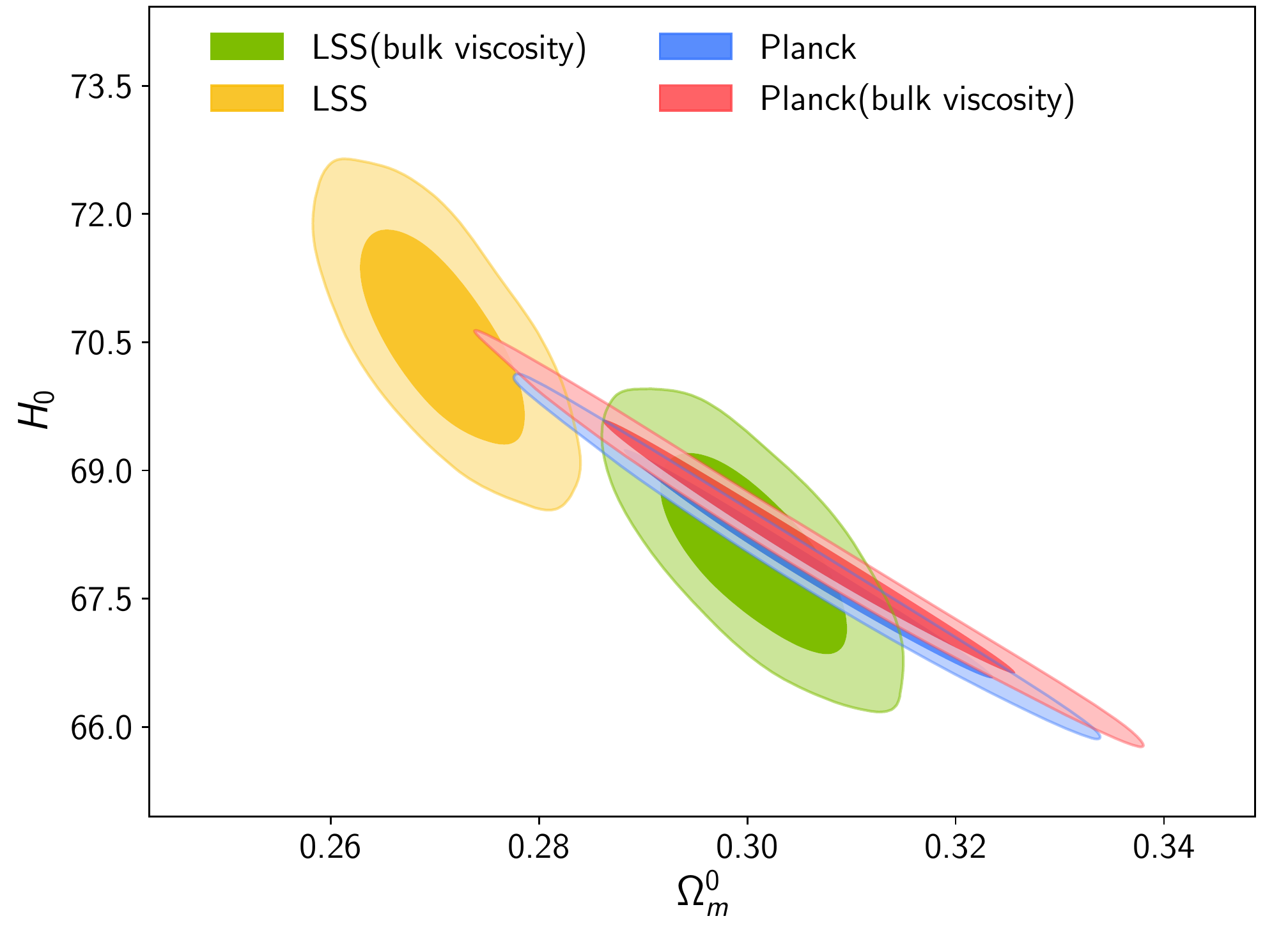}}
\hspace{-0.4cm}
\subfloat[\label{fig:H0-om-shear}]
{ \includegraphics[scale=0.40]{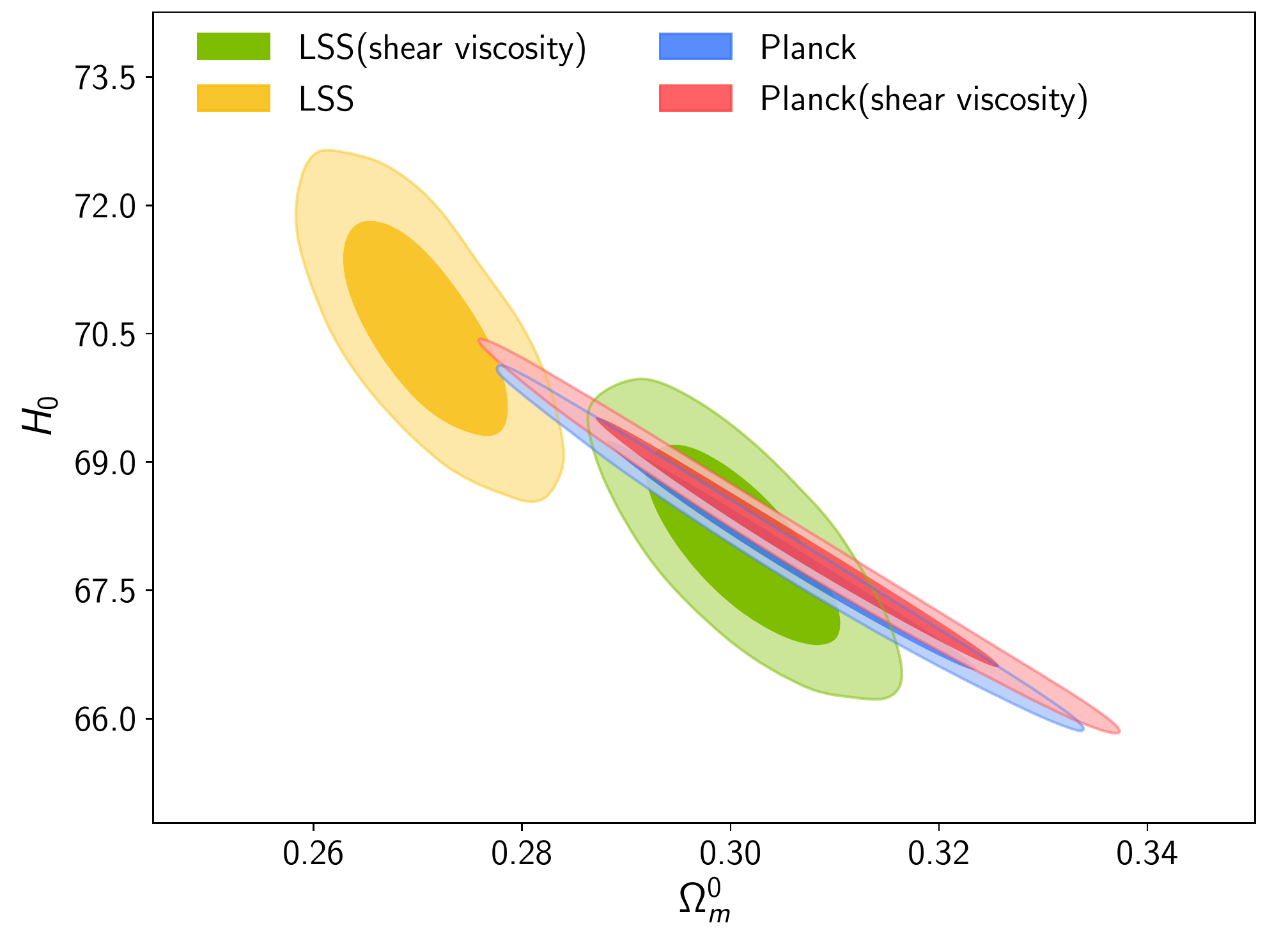}}
  \caption{Bestfit parameter range for $H_0-\Omega_m$ shows a clear discordance in $1\sigma$ level. This discordance gets resolved
  by introduction of (a)bulk viscosity or (b) shear viscosity in the CDM.}
  \end{figure}
Measurements of the value of Hubble parameter are done in two different ways. One is the direct measurement from type-IA supernova
and another one is the indirect estimation through LSS and CMB observations. 
From LSS observations $H_0$ is estimated as the required damping term in the growth of the over-densities, while 
from CMB observations $H_0$ is inferred from the scale $\Theta_{\rm MC}$ of baryon acoustic oscillation. 

The tension between the direct and indirect measurements is well known in
the literature~\cite{Verde:2014qea}. However, there was no disagreement between indirect measurements until the advent of Planck data. 
The WMAP 7 year result~\cite{Komatsu:2010fb} has given such values of $H_0-\Omega_m$ which can accommodate the LSS results. 
On the other hand, MCMC analysis done with Planck data, as described in the previous section
shows some tension with LSS result obtained similarly with the Planck prior set on $\{A_s, n_s, \tau_{\rm reion}\}$.
Planck analysis gives the value of $H_0$ to be $67.91\pm 0.89$ with $\Omega_m = 0.305\pm 0.012$
where as the best-fitted value for $H_0$ turns out be $70.58\pm 0.83$ and $\Omega_m$ is $0.270\pm 0.005$
for LSS data at 1-$\sigma$ level. 
We see that there is clear discordance between these best-fit parameter regions in \fig{fig:H0-om-bulk} and \fig{fig:H0-om-shear}.

In viscous cosmology both Hubble and viscosities play similar kind of role. Hubble acts as an over-all damping term in \eqn{eq:theta}
and $\tilde\eta$, $\tilde\zeta$ go with scale dependent damping terms. Therefore on small scales viscosity compensates
the effect of Hubble reducing the value of $H_0$. On the other hand, the value of $H_0$ 
estimated from CMB does not get affected much as it is inferred from the acoustic scale
$\Theta_{\rm MC}$.
The role of viscosity in changing $\Omega_m^0$ is not straightforward. However, it can be understood from
\eqn{eq:theta} and \eqn{eq:delta-prime} that $\Omega_{\rm cdm}$ or $\Omega_{\rm cdm}^0$ comes in the denominator 
in the term with viscous parameters. Therefore, 
introduction of viscosity drags the value of $\Omega_{\rm cdm}^0$ towards higher values to compensate the effect
of viscous term. 

The MCMC analysis done in four steps, which had been described in last section, shows these shifts in
$H_0-\Omega_m^0$ plane.  This shift ensures that the discordance between LSS and CMB observations disappear. 

\section{Planck-LSS combined viscous-cosmological parameters}
\label{sec:combined}
%===================================================================
We have seen that bulk and shear viscosity removes the tension between Planck and LSS 
observations. Therefore, we proceed to do a joint MCMC analysis using these two data sets. 
We keep bulk and shear viscosity parameters, $\tilde\zeta$ and $\tilde\eta$, varying to find the
best-fit value for them; then in two different analyses, we kept one of the viscosity parameters
to be zero to get the best-fit value of other. Since bulk and shear viscosities play almost similar 
role the best-fit value of shear(bulk) in the absence of bulk(shear) viscosity is different 
than that of the combined analysis. 

We do not find any significant change in all the cosmological parameters
from the Planck-fitted results. 
The only significant shift is visible in the derived parameter $\sigma_8$ which settles down to a lower
value than the Planck-fitted value. The value of the newly fitted parameters are shown in table-\ref{tab:para}.
The best fit value of $\sigma_8$ obtained from the analysis done with either type of viscosity 
does not change significantly from that of the bulk-shear combined analysis.

\begin{table}
\centering
 \begin{tabular} {| l  c c |}
\hline
 Parameter &  1-$\sigma$ value & 2-$\sigma$ value \\
\hline \hline
{\bf Cosmological parameters} & & \\
{$\Omega_{b }h^2$} & $0.0222\pm 0.0002         $ & $0.0222\pm 0.0004   $ \\ 

{$\Omega_{\rm cdm }h^2  $} & $0.1185\pm 0.0012          $ & $0.1185^{+0.0024}_{-0.0024}$\\

{$100\Theta_{\rm MC} $} & $1.04212\pm 0.00039        $ & $1.04212^{+0.00076}_{-0.00077}$\\

{$\ln( 10^{10} A_{s })$} & $3.070\pm 0.023            $ & $3.070^{+0.044}_{-0.045}   $\\

{$n_{s }         $} & $0.9674\pm 0.0043          $ & $0.9674^{+0.0086}_{-0.0084}$\\

{$\tau_{\rm reion }   $} & $0.069\pm 0.012            $ & $0.069^{+0.023}_{-0.024}   $\\
\hline
{\bf Viscosity parameters} & & \\
{$\tilde{\eta}$} & $1.20^{+0.40}_{-1.00}\times 10^{-6} $ & $1.20^{+1.00}_{-1.00}  \times 10^{-6}$\\

{$\tilde{\zeta}$} & $1.32^{+0.50}_{-1.00} \times 10^{-6} $ & $1.32^{+2.00}_{-1.00} \times 10^{-6}$\\
\hline
In absence of bulk viscosity & &
\\ {$\tilde\eta$}& $2.29^{+0.50}_{-0.60}\times 10^{-6}           $&  $2.29^{+1.00}_{-1.00}  \times 10^{-6} $\\

\hline
In absence of shear viscosity & & \\
{$\tilde\zeta$} & $2.46^{+0.50}_{-0.60} \times 10^{-6}           $ & $2.46^{+1.00}_{-1.00} \times 10^{-6}$\\
\hline
Derived parameters & & \\
{\boldmath$H_0$} (Km/sec/Mpc) & $68.39\pm 0.56             $ & $68.4^{+1.1}_{-1.1}        $\\

{\boldmath$\sigma_8        $} & $0.754\pm 0.011            $ & $0.754^{+0.022}_{-0.021}   $\\
\hline \hline
\end{tabular}
\caption{Best-fit values of cosmological parameters along with the viscosity parameters 
and the derived parameters for viscous cosmology are shown here. These values are obtained 
from Plank-LSS joint analyses in the presence
of both bulk and shear viscosities. These values remain almost unchanged for the analyses
with only one type of viscosity.}
\label{tab:para}
\end{table}

\section{Discussion and Conclusion}
\label{sec:conc}
%===========================================================
Through out this paper we have discussed the effect of two different viscosities on large scale 
structure formations and CMB. We have found that either of the two viscosities or their combination 
affects the growth of linear overdensity which in turn changes the matter power spectrum at small
length scales. Motivated by this, we move on to quantify the amount of viscosity supported 
by cosmological observations. Therefore, we consider the viscosity coefficients as model
parameters and perform MCMC analysis with Planck and LSS data. In the analyses with LSS data, 
the values of amplitude of primordial perturbations
and scalar spectral index is set to be equal to the value obtained from Planck CMB analysis.
We find that some amount of viscosity is preferred by LSS observations. Most interestingly 
this bestfit value of viscosity resolves the conflict between Planck CMB and LSS observations, 
both in $\sigma_8-\Omega_m^0$ plane and $H_0-\Omega_m^0$ plane, simultaneously. It is 
interesting to note that the value of $H_0$ inferred from Planck does not change 
significantly due to the viscosities, while the same obtained from LSS changes significantly. 
This is due to the following reasons: $H_0$ is obtained from the baryon acoustic oscillation scale 
and depends on the value of $\Omega_m^0$~\cite{Bassett:2009mm}.
The LSS experiments constrain $\sigma_8$
and $\Omega_m^0$ jointly~\cite{Kilbinger:2012qz} which gives a scope to accommodate lower 
$\sigma_8$ by increasing $\Omega_m^0$. However, in the case of Planck data, $\sigma_8$ is a derived parameter
which comes down to a lower value, due to inclusion of viscosity, without affecting $\Omega_m^0$.
Therefore, in the case of $\sigma_8$, both Planck and LSS fitted values change on
inclusion of viscosities, but for the $H_0$ only LSS value gets affected.

We find that the required value of bulk and shear viscosity parameters ($\tilde\zeta$ and $\tilde\eta$) ,
as obtained from MCMC analyses,
are of same order (${\cal O}(10^{-6})$) and have similar effects.
It is almost impossible to distinguish the effects arising from these two viscosities. 
The best-fit values for commonly used viscosity parameters $\eta$ and 
$\zeta \sim 3\times 10^2 $ Pa.sec\, . 

The origin of these viscosities can have different sources. The fundamental viscosity generated
by the self-interaction between the dark matter particles might be one source. Another source 
can be the large-scale integrated effect of the small scale non-linear gravitational
phenomenon~\cite{Blas:2015tla}. One might also attribute this kind of viscosity to the
corrections in the gravity sector of the Einstein-Hilbert action~\cite{Abebe:2011ry}.

The fundamental viscosity generated
by the self-interaction between the dark matter particles can be written in terms of 
the cross-section to mass ratio, $\sigma/ m$~\cite{Gavin:1985ph}. However it depends 
on the details of the decoupling history of dark matter. So the relation between
the viscous coefficients $\eta,\, \zeta$ and $\sigma/ m$ is model dependent.
The observational bound from bullet-cluster~\cite{Kahlhoefer:2013dca}
on the cross-section to mass ratio $\sigma/ m$ for self-interacting dark matter is 
$\sigma/ m \leq 1.2 \,{\rm cm^2/g}$. This gives the value of mean free path of the dark matter
particle greater than the horizon size at present day.
Therefore a hydrodynamic description might not be valid during late time, but 
it can work in the past when mean free path was within the horizon.

Previous attempts in the literature to remove the tension between LSS and Planck CMB 
for $\sigma_8$ and $H_0$ either include sterile neutrinos or exotic interaction in dark sectors.
However, these attempts fails to resolve both the discordances simultaneously. 
We, on the other hand, did not introduce any 
extra matter component to the $\Lambda$CDM cosmology. Moreover, we solve the two issues of
 $\sigma_8$ and $H_0$ with introduction of only one parameter, either bulk or shear viscosity.
 The origin of these dissipative effects requires a thorough investigation of the properties
 of dark matter in future.

 \section*{Acknowledgement:} We would like to thank Thejs Brinckmann and Miguel Zumalac\`arregui
 for their help related to MontePython.

% \subsection{Bound from SDSS and BAO}
% 
% \subsection{Bound including Lyman-$\alpha$}
% From the observed flux in the Lyman-$\alpha$ forest matter power spectrum had been reconstructed in \

%\newpage`
\bibliographystyle{JHEP}%unsrt
\bibliography{viscous-cdm.bib}  
  
\end{document}